\documentclass[sigplan,screen]{acmart}
\AtBeginDocument{%
  }

\setcopyright{cc}
\setcctype{by-nc}
\acmDOI{10.1145/3759429.3762628}
\acmYear{2025}
\copyrightyear{2025}
\acmISBN{979-8-4007-2151-9/25/10}
\acmConference[Onward! '25]{Proceedings of the 2025 ACM SIGPLAN International Symposium on New Ideas, New Paradigms, and Reflections on Programming and Software}{October 12--18, 2025}{Singapore, Singapore}
\acmBooktitle{Proceedings of the 2025 ACM SIGPLAN International Symposium on New Ideas, New Paradigms, and Reflections on Programming and Software (Onward! '25), October 12--18, 2025, Singapore, Singapore}
\received{2025-04-25}
\received[accepted]{2025-08-11}

\usepackage{listings}
\usepackage{microtype}
\begin{document}

% \title{What You See Is What It Does:\\A Structural Pattern for Legible Software}

% \author{Eagon Meng}
% \email{eagon@mit.edu}
% % \orcid{1234-5678-9012}
% \affiliation{%
%   \institution{MIT CSAIL}
%   \city{Cambridge}
%   \state{MA}

% \author{Daniel Jackson}
% \email{dnj@mit.edu}
% % \orcid{1234-5678-9012}
% \affiliation{%
%   \institution{MIT CSAIL}
%   \city{Cambridge}
%   \state{MA}
\title{What You See Is What It Does:\\A Structural Pattern for Legible Software}

\author{Eagon Meng}
\orcid{0009-0004-0855-1584}
\affiliation{%
  \institution{Massachusetts Institute of Technology}
  \city{Cambridge}
  \country{USA}
}
\email{eagon@mit.edu}

\author{Daniel Jackson}
\orcid{0000-0003-4864-078X}
\affiliation{%
  \institution{Massachusetts Institute of Technology}
  \city{Cambridge}
  \country{USA}
}
\email{dnj@mit.edu}

% \renewcommand{\shortauthors}{}

\begin{abstract}

The opportunities offered by LLM coders (and their current limitations) demand a reevaluation of how software is structured. Software today is often ``illegible''---lacking a direct correspondence between code and observed behavior---and insufficiently modular, leading to a failure of three key requirements of robust coding: incrementality (the ability to deliver small increments by making localized changes), integrity (avoiding breaking prior increments) and transparency (making clear what has changed at build time, and what actions have happened at runtime).

A new structural pattern offers improved legibility and modularity. Its elements are concepts and synchronizations: fully independent services and event-based rules that mediate between them. A domain-specific language for synchronizations allows behavioral features to be expressed in a granular and declarative way (and thus readily generated by an LLM). A case study of the RealWorld benchmark is used to illustrate and evaluate the approach.
\end{abstract}



\maketitle

\section{Introduction}
Reactions to the growing use of large language models (LLMs) in programming have focused on their promise to automate coding tasks and eliminate drudgery. Less attention has been paid to the respects in which LLMs have exposed deep flaws in the practice of software development, and how a reevaluation might be needed to capitalize on the benefits of LLMs and mitigate their failings. This paper suggests that such a reevaluation might improve the experience of programming not only for LLM coders but for human coders alike.

Most obviously, LLMs overturn a central premise of agile development---that developers should focus on code and that non-code artifacts are of suspect value. Specifications are back, even if they are now called ``prompts.''

Less obviously, modularity is ripe for a return. The extraordinary human capacity to tolerate complexity, even when self-inflicted, allows projects to erode (or never clearly define) module boundaries. A lack of investment in structure that eventually exacts a heavy cost is often justified as ``technical debt.'' When well-defined modules do exist, they may correspond only weakly to the functionality (that is, the observable behavior), so that the code lacks ``legibility,'' and mapping desired changes in behavior to modifications of code becomes needlessly difficult.

When an LLM is used to add code to an existing repository, it can be hard to control which modules are modified, and to ensure that existing functionality is not broken. A benchmark \cite{swebench} that measures coding performance on realistic tasks (drawn from a collection of about 2,300 GitHub issues) ranks LLMs by the proportion of coding challenges they can resolve. But a careful study of the benchmark results \cite{swebench-york} found that about a third of the resolved issues contained the solution code in the description of the issue itself; another third were deemed correct (because they passed test cases) despite being wrong; and more than 90\% had solutions available online before the training cutoff for the evaluated LLMs. When these flaws were taken into account, the success rate fell precipitously.
% : for GPT-4, for example, from 12.47\% to less than 1\%.

Programmers complain that LLM coding assistants recommend patches that often break previously generated functionality \cite{METR_2025_impact}, and ``whole app'' builders seem often to hit a brick wall, incapable of extending functionality beyond certain (undefined) limits. If LLMs are not able to work \textit{incrementally}, growing a system's functionality with new features while preserving the \textit{integrity} of existing ones, their role will be limited.

This paper presents a structural pattern designed to address these issues. Its key elements are \textit{concepts}---independent services with well-defined purposes---and \textit{synchronizations}---granular event-based rules that act as mediators between concepts, orchestrating data and control flow without introducing dependencies between them.

Improved \textit{legibility} comes from the direct correspondence between concepts and coherent units of user-facing functionality, and between synchronizations and (often application-specific) behavioral rules. \textit{Integrity} follows from the decoupling of concepts and their completeness with respect to their purposes, so that critical properties are localized and modifications of one concept do not propagate to another. Improved \textit{incrementality} follows from the independence of concepts, allowing each to be generated and modified without knowledge of others, and from the granularity of synchronizations, which often allows behavioral extensions to be expressed as the addition, deletion or replacement of small synchronizations.

Contemporary codebases often lack not only modularity but also \textit{transparency}. When executions fail, it can be hard to figure out what happened, let alone why. Concepts and synchronizations allow greater transparency because control flow never passes from one concept to another without an explicit synchronization, so there is no need to look inside concepts (or worse, resolve the bindings of arguments or state components to callbacks) to figure out which actions occurred. Moreover, the synchronization engine produces action traces as a natural byproduct of execution, recording the \textit{provenance} of each action so that auditing and problem diagnosis can be easily performed.

The contributions of this paper include a specification format for concepts (Section \ref{concept-language-section}), a language for synchronizations (Section \ref{sync-language-section}), and a design for an engine that executes synchronizations (Section \ref{architecture-section}). Together, these allow effective LLM-based generation of code for the backend of a web application. A case study was conducted (Section \ref{case-study-section}) for the RealWorld benchmark \cite{realworld} (a Medium-like blogging application that has been implemented in over a hundred different technologies), serving as an evaluation of a realistic (if small) application, and offering examples that will be used for illustration throughout the paper.

\section{Concepts}\label{concepts-section}
In his recent book \cite{eos}, Daniel Jackson proposes ``concepts'' as a way to structure the functionality of a software system. Concepts are user-facing units of functionality that have a well defined purpose and thus deliver some recognizable value. For example, the concepts of a social media app might include \textit{Post}, \textit{Comment}, \textit{Upvote}, \textit{Friend}, \textit{Password}, etc.

Together, the concepts of an application comprise its conceptual model, but importantly concepts are understandable independently of one another, and so when a concept refers to values from other concepts they are always treated as fully polymorphic. Thus, for example, in a social media app, the \textit{Comment} concept would be defined so that a comment's target can be any type of object, without any constraints or expectations, and in each particular app the targets will be concrete objects created by other concepts---for example, posts in a social media app, products in an online store, articles in a newspaper, and so on.

Concepts are often familiar to users, since the same concepts arise in (near) identical form in many different contexts. Indeed, this is what makes many applications readily usable, with the user bringing an understanding of most of the concepts from prior experiences with those same concepts in other applications.

In contrast to the way the word is sometimes used in other settings, a concept is not an element in an ontology. In the field of conceptual modeling, a ``conceptual model'' is often a data model in which the concepts are entities. Richer kinds of conceptual model have been defined that incorporate behavior, for example, by defining the concepts as objects (or more accurately classes) with their own internal behavior.

These approaches do not allow a modular treatment of concepts, however. The behavior associated with a concept typically involves multiple classes of objects, and modifications that involve their relationships to one another. For example, the behavior of the \textit{Upvote} concept is to associate votes with both particular items and with the users who issued the votes (in order to prevent double voting).

In Jackson's scheme, a concept is defined as its own state machine, with actions that can be performed by the user and a visible state. Thus the actions of the \textit{Upvote} concept might be to \textit{upvote}, \textit{downvote} or \textit{unvote}, each taking an item and a user as arguments; the state would include which votes were issued by which users, for which items, and whether they were upvotes or downvotes.

This idea of modeling behavior abstractly as a state machine is conventional, and central to almost all formal specification languages, including Alloy, B, VDM and Z. What is new is the separation of concerns that concepts embody, in which reusable facets of behavior are teased apart. For example, in contrast to a traditional object-oriented approach in which a \textit{User} class might include  usernames and passwords (for authentication), display names (for profiles), email addresses (for notification), and so on, in a concept design these would be factored into distinct concepts. There are many precedents to this idea, in view specification \cite{jackson-z-views} and in aspect-oriented \cite{aop}, role-oriented \cite{ooram} and subject-oriented programming \cite{sop}. 

When concepts are composed into a system, the traces (behavioral histories) of the system as a whole are always interleavings of the traces of the individual concepts. The properties of the concepts---as witnessed by the traces they permit---are thus preserved by definition. This means that many (but not all) of system properties can be localized within concepts. For example, the \textit{Password} concept will guarantee that a user attempting to execute an authenticate action will succeed only when the presented password matches the one submitted earlier on registration. Whether actions of other concepts are appropriately authenticated, on the other hand, will depend on how those concepts are composed, so the property that a user cannot delete a post unless authenticated as its author, for example, will necessarily involve multiple concepts (eg, \textit{Password} and \textit{Post}) and the way in which they composed.

In his book \cite{eos}, Jackson argued that concepts should structure not only the functionality visible to users but also the underlying implementation. A concept is thus at the same time both a protocol describing a pattern of human behavior and a service supporting it. In respect of implementation, concepts are thus similar to microservices, but whereas microservices can call each other and even query each others' state (often resulting in a tangled web of connections), concepts have no such dependencies. Nevertheless, a concept may have dependencies on lower level services, such as databases, networking services and datatype libraries.

\section{Synchronizations}\label{syncs-section}

Concepts are orchestrated by an application layer that makes calls to concept actions and queries concept states in response to incoming requests. All data and control flow between concepts is expressed in the code of this layer. This is how coupling between the concepts themselves is avoided, allowing concepts to be designed independently and then composed later into applications. 

One possible implementation pattern would be to write a procedure for each application endpoint that makes calls and queries to multiple concepts, assembles the resulting data, and then returns a response to the client. This is perhaps the simplest approach to understand and it corresponds most closely to standard practice in web applications. Indeed, this is how students in a senior-level software engineering class in the authors' university have been taught in previous years to realize concept design in their projects.

In his book \cite{eos}, Jackson proposed a more granular mechanism for composing concepts, in which a set of concepts is coordinated with a set of \textit{synchronizations}. Each synchronization has a firing condition that corresponds to the execution of a concept action, and causes the execution of one or more additional concept actions. The entire sequence of actions that results is required to form a transaction, so that if any action should fail, all the actions must be undone.

This scheme supports not only simple causal relationships (for example when a user comments on a post, a notification is sent to the post's author) but also allows actions to be used for suppression. A synchronization might say that downvoting of a post through the action \textit{Upvote.downvote(u,p)}\footnote{The concept is called \textit{Upvote} since its purpose is to allow users to upvote items, but confusingly it also includes a \textit{downvote} action.} leads to the execution of an action \textit{Karma.permit (u, 10)} representing the fact that user \textit{u} is permitted to perform some (other) action requiring 10 karma points. If the user lacks the requisite karma points, that action fails, and so the original downvoting action will be aborted also.

The granularity of this mechanism supports a straightforward mapping of behavioral properties to synchronizations. There might be a separate synchronization, for example, for each of the following:

\begin{itemize}
    \item Authentication: when a post is deleted, the user must be authenticated;
    \item Moderation: when a moderator rejects a post, the post is deleted;
    \item Cascade: when a post is deleted, its associated comments are deleted too;
    \item Notification 1: when a moderator rejects a post, its author is notified;
    \item Notification 2: when a comment is deleted, its author is notified.
\end{itemize}

Together, these synchronizations might define a complex sequence of actions: for example, that a moderator rejects a post, causing (a) the moderator to be authenticated, (b) the author of the post to be notified, (c) the post to be deleted, (d) the comments on the post to be deleted too, and (e) the authors of those comments to be notified. 
The granular structure makes it easier to enforce behavioral rules consistently, ensuring, for example, that comments on a post are deleted not only when the post is deleted explicitly by its author but also when it is deleted due to moderation.

Despite its advantages, this scheme has deficiencies:

\begin{itemize}
    \item The causal model is not easy to understand, because it mixes a CSP-like \cite{csp} symmetrical rendezvous with a directed stimulus/response pattern.
    \item Transactions complicate the implementation, making it hard to implement actions, such as notifications, whose effect in the world must be delayed until the transaction commits.
    \item Extending synchronizations to fire on multiple actions, or to condition on concept state in rich ways, is challenging.
    \item It is not clear how to accommodate error handling.
    \item Synchronizations that require actions to be mapped over a set (for example, a cascading delete that should be applied to all comments associated with a deleted post) are not easy to specify.
\end{itemize}

% To enable the scheme to be used in student projects without a transaction mechanism, a simple discipline was proposed: the actions in a synchronization would be ordered so that an action that is expected to fail under common circumstances (for example, an authentication) is placed before any actions that perform side-effects. A transaction could then be aborted by simply terminating the sequence of actions prematurely, without the need for any rollbacks. This proved workable but inconvenient, and placed unreasonable constraints on the design of actions and synchronizations, so it was abandoned.

A new synchronization scheme was therefore devised and is a key contribution of this paper. This scheme, described below, is more expressive, easier to understand and requires no transactions, all while preserving (and in some respects improving) the ability to write granular synchronizations. The elements of this scheme are as follows:

\begin{itemize}
    \item Synchronizations follow the simple causal structure of traditional event-based systems: when some actions occur under some conditions, some other actions are invoked.
    \item The firing condition of a synchronization can include any number of action occurrences, along with conditions on the states of any number of concepts, and when the rule fires, multiple actions can be invoked.
    \item By using free and bound variables that are scoped across the entire rule and implicitly quantified, firing can be predicated on concept states (for example, firing only on deletions of a post by its author), and actions can easily be invoked over a set (for example, deleting all the comments with a particular post as a target).
\end{itemize}

This scheme eliminates the need for transactional semantics, while still allowing synchronizations to be specified that fire on action failures (that is, actions that return errors or raise exceptions), thus allowing errors to be handled without special treatment.

A key insight that made this possible was to regard the incoming request as itself an action to which a synchronization reacts. A kind of ``bootstrap concept'' is introduced to model the user as the entry point of interaction, with a \textit{request} action for an incoming request and a \textit{response} action for a response. 

Now, authentication and authorization can be handled in a more natural way. For example, to say that an \textit{downvote} action can occur only when the user has some number of karma points, the synchronization would say "when a \textit{request} is received for a \textit{downvote} for item \textit{i} by user \textit{u}, where user \textit{u} has at least\textit{N} karma points, then the \textit{downvote} action by \textit{u} on \textit{i} is invoked." 

% Treating web requests and responses as actions of a pseudo concept has another happy consequence: that concept encapsulates all HTTP-specific functionality, and removes the need for the top-level procedures commonly known as "routes" or "endpoints." In web frameworks, these procedures typically rely on fancy language features (such as decorators), and mix together different concerns (URL and request parsing, request and response handling, and invocation of backend services). Eliminating them simplifies the application layer and factors out platform-dependent details.

\section{Concept Specifications}\label{concept-language-section}

% A concept specification plays two roles: as a requirement for generating and evaluating the concept's code, and as an interface for how synchronizations access the concept.

% Synchronizations orchestrate concept behaviors by calling concept actions and querying concept states, and their design must therefore depend on knowledge of concept interfaces, comprising for each concept the form and semantics of its actions and the schema of the data model of the state. In other words, synchronizations are dependent on the specifications of concepts.

% A specification language for concepts is therefore essential. This language can be rather minimal, and formal only to the extent that it provides the type signatures of actions and state components. 

Our specification language is mostly informal, except for the definition of state components and action signatures. Here, for example, is a specification for a \textit{Password} authentication concept.
% \footnote{The full specification appears in Appendix \ref{concept-spec-appendix}}: 

\begin{small}
\begin{verbatim}
concept Password [U]
purpose 
    to securely store and validate user credentials
state
    password: U -> string    
    salt: U -> string        
actions
    set [ user: U ; password: string ] 
        => [ user: U ]
        generate a random salt for the user
        compute a hash of the password with the salt
        store the hash as password and salt in salt
        return the user reference
    set [ user: U ; password: string ] 
        => [ error: string ]
        if password does not meet requirements
        return the error description
    check [ user: U ; password: string ] 
        => [ valid: boolean ]
        retrieve salt for user
        compute hash of provided password with salt
        compare with stored hash
        return true if hashes match, false otherwise
    check [ user: U ; password: string ] 
        => [ error: string ]
        if user does not exist or has no password
        return the error description
    validate [ password: string ] 
        => [ valid: boolean ]
        check that the password meets requirements
        return true if valid, false otherwise
operational principle
    after set [ user: x ; password: "secret" ] 
        => [ user: x ]
    then check [ user: x ; password: "secret" ] 
        => [ valid: true ]
    and check [ user: x ; password: "wrong" ] 
        => [ valid: false ]
\end{verbatim}
\end{small}

The specification is broken into named sections according to the structure in Jackson's book \cite{eos}. The state section specifies the concept's data model as a collection of relational state components in the style of Alloy \cite{alloy}. The \textit{password} and \textit{salt} components, for example, associate a hashed password and a salt value with each user. Note that users are represented by a type variable \textit{U} which is listed at the top as a parameter of the concept. Type parameters cannot be constrained, so this ensures no external coupling or dependence on whatever type is actually used to represent users.

The action section lists each of the concept's actions with their signatures and informal specifications. Each action is split into cases in the pattern-matching style of functional programming languages; in this example, the input arguments always take the same form, and the cases happen to be split only on the output arguments. Each action can take any number of arguments and return any number of results. The names of the arguments and results are significant. This will be important both to allow matching in synchronization rules on only a subset of arguments, and also to distinguish normal and exceptional outcomes. The \textit{set} action, for example, which sets the password of a user, takes (a reference to) the user and a string representing the password, and returns either the same user (when the action is successful) or an error message (otherwise). Note that the specification of the action is only partial; it does not enumerate the well-formedness rules for passwords, for example, because these details are not needed for synchronizations.

The operational principle is a kind of archetypal scenario that describes how the concept fulfills its purpose, and plays a useful role both in understanding the concept and in generating test cases.

Concept specifications can play a variety of roles. Most importantly, they define the interfaces to synchronizations. They can be extracted from concept code, but can also be used to generate concept code.  They can be written by human designers, or generated by an LLM. In this case, the \textit{Password} concept was generated by an LLM in response to a minimal prompt (see Section \ref{developing-experience}), and was then used to generate the concept code as well. The prompt asked only for a standard password concept, but the LLM included details of hashing and salting in both the specification and the generated implementation. Two other concepts (see Appendix \ref{concept-spec-appendix}) that involve users were defined for the case study: a \textit{User} concept that allows users to be identified by email addresses or usernames, and a \textit{Profile} concept that associates with users information for public display.

Concepts, unlike abstract data types, expose their states to the application layer, albeit in an abstract form. This is important for three reasons:
\begin{itemize}
\item It saves designers and programmers from the trouble of having to anticipate (and then deliver on) all the different ways in which the state may be queried.

\item It allows synchronizations to use database queries that cross concept boundaries---for example, delivering the bio from a user's profile given their username, joining the relations of the \textit{User} and \textit{Profile} concepts---which makes code simpler, more uniform and more efficient.

\item It preserves the idea that concept behaviors are user-facing, which treating state observations as actions would compromise. This does not mean that all actions have to update state; the check action of the \textit{Password} concept is genuinely an action that would make sense to a user (in contrast, for example, to an observer action that merely obtains a reference to a user from the user's email address, a query that synchronizations might perform for reasons that are of no interest to the user).
\end{itemize}

\section{Synchronization Language}\label{sync-language-section}

Synchronizations describe how the actions of independently defined concepts interact, and have the standard form:
% They represent the natural grammatical structure for describing behavior of: 
``\textit{when} these actions happen, \textit{where} the current state of things are so, \textit{then} these actions should follow.'' The examples below illustrate the expressivity and legibility of this form.

\subsection{A Basic Synchronization}\label{basic-syncs}
Our first example illustrates the basic ideas and structure of a synchronization for a simple user registration flow:

\begin{small}
\begin{verbatim}
sync Registration
when {
    Web/request: [
        method: "register" ;
        username: ?username ; 
        email: ?email ] 
        => [] }
where { bind ( uuid() as ?user) }
then {
    User/register: [
        user: ?user ; 
        name: ?username ; 
        email: ?email ] }
\end{verbatim}
\end{small}

This says: when a \verb|Web| request for \verb|register| occurs for some \verb|?username| and \verb|?email|,  \verb|bind| a new \verb|uuid()| as \verb|?user|, and \verb|register| a new user with that \verb|?user| identifier, and with a \verb|name| of \verb|?username| and an \verb|email| of \verb|?email|. 

The parts (all required except for \verb|where|) are:
\begin{itemize}
    \item \verb|sync|: a unique name for the synchronization hinting at its purpose;
    \item \verb|when|: a pattern for matching action \textit{completions} (actions that have returned with an output, denoted by the \verb|=>| pointing to a pattern for the output record);
    \item \verb|where|: a query on the states of one or more concepts, as well as any calculations not requiring state;
    \item \verb|then|: a list of action \textit{invocations} to be executed by their associated concepts (action patterns without a \verb|=>|).
\end{itemize}

Clauses may include (in syntax similar to SPARQL \cite{sparql}):

%, and allows keywords like \verb|bind| and \verb|filter| for further expressivity if needed. In general, the terms featured are:

\begin{itemize}
    \item Literals (such as strings and numbers);
    % \verb|"literals"| \textemdash{} standard value terms like "strings", 3240 (numbers), etc.
    \item Variables (with a question mark)
    that match action arguments or state components;
    % \verb|?variables| \textemdash{} declarative bindings that declare that joins two different actions or their concepts' state
    \item Namespaces (such as \verb|Concept:| for a concept, or\\
    \verb|Concept/action:| for an action) that fully qualify all property names within;
    \item Properties (\verb|property:|) which appear within brackets, and are \textit{context-specific} and qualified by the surrounding namespace. Inside a \verb|Concept/action:| namespace, these refer to the input and output arguments of that action; inside a \verb|Concept:| namespace, (illustrated in the where clause of examples below) these refer to the state relations of that concept.
\end{itemize}

% Comments can also be included, where only the \verb|#| character for indicating everything until the new line is a comment, is supported.

% Importantly, all such action signatures in the `when` clause must include some output as in \verb|=> [ output: ?output ]|, where the \verb|=>| refers to the fact that this is an action \textbf{completion}, and the record specified is used for any potential synchronizations against the output of the action. As shown here, you do not need to specify any specific output if you are not interested in it, but you do need to specify that there \textit{was} an output, represented by the \verb|[]|.

\subsection{Granularity and Partial Matching}

Actions in the \verb|when| clause can match solely based on the specified arguments, allowing for enhanced granularity. After a \verb|User| registration, the following synchronization handles setting a \verb|Password|:

\begin{small}
\begin{verbatim}
sync NewPassword
when { 
    Web/request: [ 
        method: "register" ; 
        password: ?password ] 
        => []
    User/register: [] 
        => [ user: ?user ] }
then { 
    Password/set: [ 
        user: ?user ; 
        password: ?password ] }
\end{verbatim}
\end{small}

Notice that the input pattern for \verb|User/register| is empty, indicating that for the purposes of this synchronization, any successful registration (denoted by the \verb|user:| argument present in the output; see Appendix \ref{concept-spec-user} for the full concept specification) is grounds for setting a password. The other condition is a \verb|Web/request| for the \verb|"register"| method, which could contain more arguments---as seen in the previous section---but which do not need to be specified here.

% This partial matching not only aids clarity and brevity, but helps reuse of synchronizations across applications with varying levels of specificity in arguments.

\subsection{Errors and Flows} \label{error-and-flows}

% The next example demonstrates how error handling occurs:
Error handling involves matching on outputs that are errors:

\begin{small}
% # User registration error is a validation error
\begin{verbatim}
sync RegistrationError
when {
    Web/request: [] 
        => [ request: ?request ]
    User/register: [] 
        => [ error: ?error ] }
then { 
    Web/respond: [
        request: ?request ; 
        error: ?error ; 
        code: 422 ] }
\end{verbatim}
\end{small}

% This example demonstrates another useful feature of the synchronization language: the use of argument names to match on just a part of an action input or output. As seen above, the \verb|Web/request| action can have other arguments, but in this error case they are irrelevant. If the \verb|User/register| action invoked by the previous synchronization fails, this synchronization will match on the error and invoke a response describing it. Note that in the previous example, the empty result list limits the match to a completed action, but allows any result.

This example demonstrates how error handling can be modeled without special features or language extensions. The \verb|error:| argument is simply another name for the purposes of pattern matching, and is fully defined in one of the action overloads in the concept specification. Different kinds of exception or result are thus easily defined.

% Note that we have now additionally modeled and are binding against the output of \verb|Web/request|, in particular its \verb|request: ?request|, so that we can respond specifically to that request. The idea of using the key \verb|error:| for \verb|User/register| is not arbitrary, and is again behavior described explicitly as one of the transitions in the concept's actions. We simply leverage that behavior in the synchronization to drive other actions conditionally.

%\subsection{Flow: Syncing Causally Related Actions}

This example also illustrates multiple actions in the \verb|when| clause. To match, such actions must have occurred \textit{in the same flow\footnote{A flow is a directed acyclic subgraph of action occurrences with its root in an external action, such as a user input or HTTP request. A more detailed treatment is given in Section \ref{flow-scope}.}}. Here, the synchronization says ``when a web request occurs, and a user registration fails, respond with an error about that failure.'' The link between the user registration failing and the specific web request, implicit in the flow, is essential, as other web requests may be in process at the same time.

\subsection{Expressing Design Decisions}

The synchronizations seen so far are unremarkable and almost inevitable. But synchronizations can also be used to express application-specific design decisions in a granular manner. For example, a synchronization may create a default profile for a user when they register successfully:

\begin{small}
\begin{verbatim}
sync DefaultProfile
when { 
    User/register: [] 
        => [ user: ?user ] }
where { bind ( uuid() as ?profile ) }
then { 
    Profile/register: [ 
        profile: ?profile ; 
        user: ?user ] }
\end{verbatim}
\end{small}
\noindent
or generate a fresh JWT token on registration, logging the user in implicitly:

\begin{small}
\begin{verbatim}
sync NewUserToken
when { 
    User/register: [] 
        => [ user: ?user ] }
then { 
    JWT/generate: [ user: ?user ] }
\end{verbatim}
\end{small}

%The first invokes the creation of a default profile for a new user when they register successfully; note the \verb|?user| property in the output of the \textit{register} action that indicates a successful result. A different application may have more than one kind of profile, or different steps for profile creation that are conditional on certain checks. In contrast, this synchronization represents a specific decision made to couple user registration to profile creation.

% (with the success being indicated by the  again that we explicitly match on the signature corresponding to a successful registration, which then contains the \verb|?user| of interest.) 

% The second synchronization generates a fresh JWT token for the user. When a user successfully registers, this token will allow them to authenticate themselves immediately. A different design might force them to go back to a login screen despite having just entered their credentials.

\subsection{Fusing Data from Multiple Concepts}

A final example shows how a synchronization can assemble a formatted response body:

\begin{small}
\begin{verbatim}
sync RegistrationResponse
when {
    Web/request: [ method: "register" ] 
        => [ request: ?request ]
    User/register: [] => [ user: ?user ]
    Profile/register: [] => [ profile: ?profile ]
    Password/set: [] => [ user: ?user ]
    JWT/generate: [] => [] }
where {
    User: { 
        ?user 
            name: ?username ; 
            email: ?email }
    Profile: { 
        ?profile 
            bio: ?bio ; 
            image: ?image }
    JWT: { ?user token: ?token } }
then {
    Web/respond: [ 
        request: ?request ;
        body: [ 
            user: [ 
                username: ?username ;
                email: ?email ;
                bio: ?bio ;
                image: ?image ;
                token: ?token ] ] ] }
\end{verbatim}
\end{small}

Most of the actions have empty argument lists, since the synchronization depends only on awaiting their successful completion and then formulating a response from the resulting state of their respective concepts.

The complexity of this synchronization (which is taken from the case study described below) reflects the ad hoc nature of the API design, and the conflation of multiple concepts in a single response. Nevertheless, it is important for synchronizations to be expressive enough for such cases.

\section{Architecture}\label{architecture-section}

The architecture comprises:
%of the system is driven by events corresponding to actions, and organized into:
\begin{itemize}
    \item Concept implementations: independent services that manage their own state, accept action \textit{invocations}, perform side effects, and respond with action \textit{completions}.
    %that signify performed actions.
    \item The synchronization engine: a service that allows synchronizations to be registered, accepts \textit{completions}, and records action \textit{invocations} for actions to be performed.
    %based on the specified synchronizations.
\end{itemize}

The synchronization engine can be thought of as a reactive database that holds action invocation and completion records, with completions triggering actions within independently running concepts. In our implementation, the concepts are implemented as modules that execute in the same thread as the engine, with the engine calling concept actions directly. Other schemes are possible: concepts might run as separate processes, and a variety of push/pull event frameworks could be used to link the engine and concepts, allowing latency and consistency to be balanced in different ways.

%This can be implemented in multiple ways, either synchronously with the engine directly calling concept actions, or asynchronously with pushes from the engine or pulls f
%with control : as a push notification from the synchronization engine to a concept, or as a pulled record polled by concepts periodically. 

%As the engine functions as a standalone database service, both can be implemented simultaneously to balance latency and consistency, with further exploration of these properties in section \ref{whole-app} describing the flow of execution for a complete application.

To deliver legibility, where ``what you see" truly is ``what it does," an implementation scheme must 
satisfy some essential properties, which we enumerate in subsequent sections along with the solutions we adopted.

\subsection{Naming}
Names are often a  source of confusion and illegibility: does \verb|user| refer to a class, an argument, or an identifier? How about names associated with different versions, environments, and applications? Each name must have a direct mapping to a unique identifier.

One approach is to use the Resource Description Framework \cite{rdf} to map names in concept specifications to Uniform Resource Identifiers (URIs). For example, the \verb|password| argument of the \verb|Password| concept's \verb|set| action might be:
\begin{small}
\begin{verbatim}
https://essenceofsoftware.com/
    concepts/0.1/Password/set/password
\end{verbatim}
\end{small}

The last three elements of the URI correspond to the three levels of names in concept specifications: concept, action and argument. Synchronizations themselves are given names too, which prove essential for auditing and debugging.

%To differentiate between concept state and actions, a period \verb|"."| is used, as in \verb|Password.salt|, to signify a relational join. 

For the case study, we used an RDF-based graph store to represent concept states, which allows predicates (relations in the concept specification) to be fully qualified names. In a more traditional setting, say of a relational database, mappings might be needed, e.g. from a column name in a table to its fully qualified name.

Argument order conventions for calling APIs can reduce legibility for unfamiliar developers or LLMs. To avoid this, we use named arguments for actions, which are easily implemented with functions that take and return a dictionary object. As explained earlier, named arguments are essential for partial pattern matching as well.

\subsection{Versioning and Causal Documentation} \label{causal}

%Documentation should not be an after-thought or ever out-of-sync with implementations. The \verb|purpose| is a canonical description of the function of the concept, and should be mapped 1-to-1 to any implementation claiming that it fulfills the concept.

Every record in the state of a running application is associated implicitly with a particular version of the concepts and synchronizations that produced it. This can be made explicit, by including in every record a reference to the application version, thus establishing a causal link between the state record and the documentation that explains it.

In our case study implementation, we leverage the built-in ability of RDF quad stores (subject, predicate, object, graph) and co-opt the graph identifier for this purpose, which allows the states of multiple versions of the running application to co-exist in the database without confusion. This approach eases transitions between staging and production environments and offers a path towards multi-tenancy (more details in Appendix \ref{state-information}).

% persistent database is used, allowing for each graph triple of all state (subject-predicate-object) to be associated with a specific versioned graph, with an identifier chosen as the specific concept specification. Each implementation of a concept is thus associated with a particular version of the concept specification and its purpose.

\subsection{Scoping via Flows} \label{flow-scope}
Traditional route handlers are often unwieldy, with control structures for handling different request variants, conditions and errors. This damages legibility and makes it hard for LLM and human coders alike to adjust or augment the behavior associated with a web request. Synchronizations reduce this complexity and improve legibility, by replacing elaborate imperative code with granular and declarative rules.

Nevertheless, the traditional routes have one advantage: that the scope of each test and action is clearly tied to a single web request. In our scheme, this scoping is implicit, and is achieved by
%What does it mean to condition on both a completed web request and a user registration error when many such events are occurring simultaneously? Inside a traditional TypeScript route handler, the code immediately following a conditional check for a user registration error is implicitly understood to be in scope for a specific web request. As such handlers
%grow in size and complexity, this reduces the legibility of any specific line of code: what the current context is, and what kinds of conditions have led to this point. 
%The idea of flow for grouping multiple conditions is designed to reify the scope, or partition, of action records to consider. Implementations for the engine must then model such scope as explicit data describing flow.
%A proposed solution is to simply 
associating a \textit{flow} token with all action records (see Appendix \ref{sync-agent} for implementation details), and propagating the token through synchronizations. All action records matched in a synchronization's \verb|when| clause must carry the same flow token, and if the synchronization succeeds, all invocations in the \verb|then| clause will be given the same token. Consequently, given an initial action, all cascading synchronizations will result in a set of causally related actions associated with the same token.

%The case study implements this as another key in all records labeled \verb|"...flow"| with the full expansion having a URI prefix associated with the specific deployment of the synchronization engine itself. A discussion of how the initial flow token is established is given in section \ref{whole-app} on what it means to be an initial or externally caused action.

\subsection{Modularity of State}
%\subsection{Modularity of State: Read/write separation and Polymorphism}
The modularity of concepts can be compromised in (at least) two ways. One is the proliferation of ``getters''
%(such as \verb|getByAuthor| for a Post concept),
driven by the query needs of clients and thus coupling concepts to synchronizations. Our solution to this is simply to have no getters. Reads and writes are strictly separated: reads are handled by client-driven querying capabilities such as GraphQL \cite{graph-ql} and writes are handled directly by the action API. In our case study, we exploited the RDF ecosystem, using a SPARQL \cite{sparql} query engine (the Comunica \cite{comunica} library) to provide federated read access to concept states.

The other is the temptation to include in the schema---of the state of a concept---references to schema components of other concepts (eg, through a foreign key). We avoid this by representing all external references with UUIDs. Unfortunately a lack of support for polymorphism in database schemas (at least in the platforms we are familiar with) means that static checking cannot yet ensure that such references are used consistently.

\subsection{Implementing the Where Clause}\label{where-section}

%Concepts maintain a greater level of modularity by not specifying a specific set of read patterns, allowing them to survive changes in application querying demands. However, this specification must still occur somewhere, and the synchronization language provides a declarative and isolated place in the \verb|where| clause to describe all reads. 

Reads and writes are separated in a concept interface, with reads implemented as database queries and writes as actions. Concepts do not query each other's states, but synchronizations can query them, often across multiple concepts (another important reason to prefer queries over getters). An implementation must therefore translate the declarative patterns of the synchronization \texttt{where} clause into whatever queries are suitable for the concept implementations.
 %and  against the generic client-driven read endpoints of concepts.
 
The \texttt{where} clause is given a uniform semantics as a function from a single binding to a set of bindings (each binding comprising values for a given set of variables). The incoming binding will hold the values of the variables bound in the \texttt{when} clause; the outgoing bindings will hold the values of variables that resulted from database queries, and will be used in the \texttt{then} clause, with one invocation of the clause for each resulting binding. For example, in a synchronization that cascades a deletion of a post to the deletion of all its comments, the query might map a single binding with a reference to the post to multiple bindings, one for each comment associated with it, allowing a simple declarative specification without needing any explicit iteration.

%A proposed solution is to treat the role of the where clause as a streamed set of partially filled bindings---from variables to matched values in the \verb|when|, and values to-be-filled from the \verb|then|---and query each concept, selecting for the missing variables given the existing bindings at the current point in the clause (as a set operation, order only matters for selectivity and efficiency).

In the case study implementation, all concept databases are RDF-compatible graph stores, and so a SPARQL \cite{sparql} engine provides this natively (given some code that expands names of concepts and actions in context to their fully qualified forms as named graph patterns).

%, which has a direct translation from the contents of the \verb|where| clause: every \verb|Concept: { ... }| structure is actually syntactic sugar for named graph patterns in SPARQL, and the parser expands the full identifier for the concept name to the current specific concept specification. In general, it should be possible to translate this kind of graph querying language to document and table-based storage by utilizing the fully-qualified names and types provided by the concept specification as a direction for future research.

\subsection{Provenance and Firing Consistency} \label{provenance-firing}

The synchronization engine can fail during the processing of synchronizations, but should be able to recover and resume execution seamlessly. To ensure both atomicity and idempotency of synchronizations, preventing duplicate invocations, an implementation must durably record synchronization evaluation itself and provide a mechanism to determine what has already occurred. 

One solution is to 
%we augment flow tokens 
augment the engine's state with \textit{synchronization edges} that connect each action occurrence back to each of the actions that precipitated it, labeled by the name of the synchronization (see Appendix \ref{sync-agent} for implementation details). A synchronization is then activated only for a particular action completion if there is no existing outgoing edge (to some invocation) from that completion, labeled with the synchronization's unique identifier.
% (within the given flow).

%the unique identifier for the synchronization is associated as the pairwise edge from each action record of the \verb|when| completions to each record of the \verb|then| invocations.

%In the case study, the engine implements this strategy by checking on every new completion whether or not the current synchronization has already been evaluated. 

This strategy allows for all records to be reevaluated on reboot during failure to resume execution. Along with flow tokens, this allows provenance tracking in which each action occurrence can be traced back both to its causal predecessors at runtime and to the synchronizations that produced it.

\subsection{Bootstrapping: The Web Concept}

For the engine scheme to work, there need to be root actions that correspond to external stimuli. These actions are treated as the actions of a special bootstrap concept. In our case study implementation, since these actions originate as HTTP requests, the \verb|Web| concept encapsulates the details of HTTP request and response structures. 

At the same time, this concept can be viewed as a proxy for actions performed in the world that connect the application to it. These actions can always be identified easily in the set of synchronizations, since they are the only ones that have completions but no invocations. This bootstrap concept is therefore the entry point to the system, and encapsulates various configuration and operational concerns.

\section{RealWorld Case Study}\label{case-study-section}

RealWorld \cite{realworld} is a benchmark for comparing platforms for web applications. It was inspired by the success of TodoMVC, a site that showcased implementations of a simple to-do app, but was limited to frontend frameworks. RealWorld specifies a common service API, and includes both frontend and backend implementations, which can (in theory at least) be tested in arbitrary combinations. The benchmark application is Conduit, a Medium-like blogging platform in which authors post articles, and readers can comment, tag, and favorite articles, follow other users, and so on. RealWorld aims to be the ``mother of all demo apps,'' and currently over 100 implementations are included.

%Because many of the RealWorld implementations were present online before the training cutoff of contemporary LLMs, the problem of generating code in any style for the RealWorld API is not especially compelling. Generating concept-structured code, on the other hand, is more interesting, because such code has not existed until now, and it requires a very different structure from existing implementations. 

The RealWorld API is only minimally documented, and consists primarily of a collection of JSON declarations specifying the format of expected responses. In concept design terms, the API conflates concerns that could be separated into distinct concepts and distinct API calls. For example, the API call to unfollow a user returns that user's profile (including their bio); similarly,  calls to obtain articles return, along with the article author's username, all the author's profile fields, whether or not the author is being followed, and how many times the article has been favorited. These complications provide a helpful challenge, since the ability to meet this spec would imply that a more cleanly organized API would be even more easily delivered.

The key goals of the case study were to determine (a) whether the benchmark app could be straightforwardly programmed using concepts and synchronizations; (b) how readily the concepts and synchronizations of the backend could be generated by an LLM; (c) and what surprises programming in this style might bring. 

%Prior to reporting on these questions, the next few sections outline the concept and synchronization languages using illustrations drawn from the case study.

\subsection{Generating Specifications and Code}\label{developing-section}

To demonstrate feasibility and expressiveness, we first built a full RealWorld backend implementation of Conduit with concepts and synchronizations written by hand, and checked that it passed the provided automated Postman test suite.

We then used an LLM to generate a complete, second version of the codebase. For each concept, we generated a specification from a brief prompt (see the Appendix for examples), and then generated code from the concept specification. Importantly, the LLM's context for the generation of both concept specifications and code is limited to the concept at hand. These generation steps almost all completed successfully in a single shot.

We then generated the synchronizations. Here the context included only concept specifications and no code. The prompts included (a) some general and rather vague requests to generate typical synchronizations for these concepts; (b) a few hints for special cases (eg, that a default profile should be created on registration); and (c) the JSON response formats of the endpoints specified in the RealWorld benchmark (to account for the idiosyncratic packaging of results).

Generating the synchronizations required some iteration. After the synchronizations were generated, we ran the Postman test suite which sometimes revealed errors suggesting additional synchronizations. We also ran some ad hoc tests which pointed to further missing synchronizations (for example, ensuring that the user deleting a comment is the author, a requirement that the standard test suite does not enforce).

Throughout this generation process, we aimed to align the LLM's knowledge of the framework with documentation intended for human consumption. To generate the synchronizations, for example, we used (an earlier version) of Section \ref{sync-language-section} as the system prompt. 
% This property proved useful during the development experience, particularly when evolving the application in the face of problems. 
This proved useful when early failures of LLM generation exposed flaws in our explanation.

% When asking for synchronizations about a subset of concepts, only those concept specifications need to be provided to guide proper usage of action signatures. In this way, both human and LLM agents share the same specification for writing synchronizations.

\subsection{Design Rules}\label{design-rules-section}

We have yet to conduct a systematic comparison of the existing benchmark implementations with ours. But an initial analysis illustrates some of the respects in which our approach may offer improved modularity and legibility.

A programmer seeking to change or extend an application's behavior relies on \textit{design rules} that confine changes within a limited scope. These design rules, often implicit, may arise from a framework that dictates the structure of the code, from common design patterns or programming styles, or from design decisions made by the programmer.

For example, most of the RealWorld implementations have three separate layers: a routing layer that manages URL and request structures; a controller layer that handles business logic; and a data or model layer that persists the data. Each layer is intended not only to separate concerns (eg, separating data usage in the controller layer from data representation in the model layer), but also to encapsulate technological dependencies (eg, on HTTP in the router layer and on the particular database in the model layer).

Each layer, moreover, is split into modules, usually along object-oriented lines, with one file for each entity. RealWorld implementations, for example, will typically have routing, controller and model modules for articles, users, tags, etc.

The design rules determine not only these organizational principles, but also constrain dependencies. Modules within a layer typically may not call each other: thus routing modules call only controller modules, which in turn only call model modules, and only model modules are expected to make database calls. Dependencies are also confined within entities: thus the router for articles is expected to call functions only in the articles controller, which is then expected to use only the articles model.

Finally, a more informal design rule maps functionality to entities. Functions associated with articles, for example, are expected to be within the article modules.

Examining a few RealWorld implementations, we found that their design rules are frequently violated\footnote{The implementations cited here are:\\
a. https://github.com/SeuRonao/realworld-express-prisma\\
b. https://github.com/gearhead041/realworld-nestjs-prisma-mongodb\\
c. https://github.com/winterrrrrff/realWorld-server}:
\begin{itemize}
    \item Controllers include database accesses that bypass the model layer [a];
    \item Controllers access the models of other controllers (article accesses user [a,b,c]; article accesses tag [a,b]; tag accesses article [b]);
    \item Routers are not always limited to their controllers (article router calls comment controller [a]).
\end{itemize}
The mapping of functionality to entities is often uncertain, because several areas of functionality are not neatly contained within entities. Different aspects of favorites, for example, are updated by functions in both the article and user schemas [a], and tagging is handled by controllers for both articles and tags [a].

Our approach, likewise, imposes design rules: 
\begin{itemize}
    \item Concept actions do not call actions or access the state of other concepts;
    \item State declarations in one concept have no dependencies on state declarations in another;
    \item Synchronizations only access concept states and actions;
    \item Only a single bootstrap concept can initiate actions, and encapsulates front-end commitments (such as the use of HTTP).
\end{itemize}
Our implementation respects these design rules by construction: the generation process isolates concepts from one another, and synchronizations from knowing the structure of the code. Synchronizations confine reads of action records to the \verb|when| clause and reads of concept state to the \verb|where| clause. The bootstrap concept is predefined and independent of any domain-specific code, so its properties are easily preserved.

The mapping of functionality to concepts is straightforward, since concepts (unlike entities) naturally accommodate relational aspects. Our implementation thus encapsulates favorite and tag related behaviors in the two corresponding concepts. In contrast, none of the standard implementations we looked at has a model for favoriting, which gets squeezed inconsistently into the user or article model, and may even appear in both [c]. Even an implementation that includes a tag model does not use it to contain all tag-related functionality [c], perhaps due to an object-oriented tendency to regard tags as properties of articles. Our implementation encapsulates translation from usernames to user ids in a single concept; one conventional implementation encapsulates it with following in a profile service [b]; one encapsulates it in a free-standing utility function [a]; and another does not encapsulate it at all [c].

\subsection{Case Study Reflections} \label{developing-experience}

On the positive side, concepts and synchronizations enabled a very granular and incremental style of development. As illustrated in Section \ref{sync-language-section}, the synchronizations for aspects of user management, profile updates, and authentication could be completed independently and step by step. Notably, each synchronization directly implemented the functionality specified, without requiring a larger scope of code (such as the rest of a ``route'') to be completed. For example, after writing the synchronizations for initiating a \verb|User/registration| when a \verb|Web/request| happens, running an HTTP request against the application results in a registration, which can be confirmed independently in the \verb|User| database, even without a synchronization that produces a web response.

While interacting with an LLM, the intermediate specification language was a useful substrate for conveying specific changes or design choices. For example, the first pass generation for a Password concept, with the generic prompt \texttt{``Create a Password concept''}, resulted in a design incorporating the additional actions of \verb|reset| and \verb|completeReset| that handled a secure password reset with a temporary reset token in the state. While this is a serviceable choice, this reset behavior can be handled in a more modular way, by synchronization with a concept like \verb|JWT|. Since the prompt for the implementation is exactly the concept design spec, simply removing these two actions and the single state component (resulting in the example in Section \ref{concept-language-section}) generated the final working implementation in a single pass. 

In general, working on top of the specification language allowed a more direct and reliable manipulation of code than issuing prompts in terms of the code would have. When working on the \verb|Profile| concept, adding a user thumbnail image was achieved within the specification: the LLM proposed to add an \verb|image| state and two \verb|update| action overloads, and generated a working version with the proper implementation. Throughout the overall process, conversational steps were mostly budgeted toward clarifications and choices about design at the specification layer, with fewer spent on implementation and generation of code.

% As all action occurrences are recorded as database records, at every point of execution the runtime history is reflected as an image of all the action records up to that point in time. Execution breakpoints become simply checkpoints of the history, a feature shared by many event-driven architectures. This was useful as 

% The separation of actions into invocations and completions provided resume capabilities in the case of malfunctioning concept implementations (in addition to the usual benefits of having full system replay and a complete audit log). For example, action records without an associated \verb|output:| can continuously await the idempotent completion of that action (with respect to the action ID) in a way similar to the transactional outbox pattern \cite{richardson2018microservices}, but used for consistency of all actions. The same system image/action history could then replay a specific invocation of an action to a concept implementation: a useful pattern to test the individual behavior of concepts situated in a real run-time context.

On the negative side, the extreme granularity of synchronizations, and the fragmentation of the conditions under which an action is invoked,
%and complete specification of behavioral and informational conditions for the invocation of any given action 
made it harder to understand the full context of a series of actions at a glance. A compensation, however, is that behavior that would be replicated across multiple endpoints in a traditional implementation was often factored into a single synchronization.

In generating specifications, LLMs can be over enthusiastic (as the password reset example above illustrates). The system prompt does not provide extensive guidance on proper concept design, and the abundance of object-oriented approaches in training material for many models today can result in designs that do not yet fully capitalize on the modularity that concepts and synchronizations make possible.

%A single synchronization is fully independent in the sense that it is declarative over the conditions it will fire under, but the combination of a series of conditions across a set of synchronizations can span nonlinear portions of a list of every synchronization. 

%These subsets can elude hierarchical grouping due to granular reuse of individual synchronizations and can complicate the story for storing them in a directory structure. 

% However, directed graphs for all relevant synchronizations are statically derivable from \verb|when|/\verb|then|, and suggest an opportunity for tooling to provide a code navigation experience that filters dynamically for all relevant parts, instead of over an arbitrary file system. The underlying capability for this tracking proved useful in the next section.

\subsection{A Bug Fixing Story}

The properties of our framework are illustrated not only when generation succeeds (especially in the dramatic reduction of the LLM context that it enables) but also when it fails. 

The following example shows how we exploited the provenance properties in debugging a faulty synchronization:

% The following section describes a unique debugging flow that takes advantage of the properties of the approach. In particular, we exploited the 
% For example, as all actions are fully reified as data in a graph, we have
% \textbf{provenance} properties:

\begin{itemize}
    \item By querying the graph of executed actions for a single flow, we were able to see the causal links between actions and trace back to the initiating action.
    \item Since all action occurrences are labeled by the synchronization responsible for them, we were easily able to find the code responsible for undesirable actions.
\end{itemize}

%Since all action invocations are labeled directly with the unique identifier of the responsible synchronization, any behavior caused by undesirable actions can be traced directly to the synchronizations that caused it.

\begin{comment}
\begin{itemize}
    \item \textbf{Runtime behavior and execution flow.} By querying the action graph using any engine of choice, and binding on the same \verb|flow|, actions can be traced to a starting action that first began the flow.
    % , such as the \verb|Web/request| described previously.
    \item \textbf{Identifying the responsible code.} Since all action invocations are labeled directly with the unique identifier of the responsible synchronization, any behavior caused by undesirable actions can be traced directly to the synchronizations that caused it.
\end{itemize}
\end{comment}

%More importantly, the actions are labeled with a semantically meaningful and unique synchronization name, allowing both human and LLM agents to share a vocabulary about how the system evolves. 

The bug in question involves the user registration flow presented in Section \ref{case-study-section}, which actually exhibits an undesirable behavior. After a user attempts a registration with a bad password, a subsequent registration attempt with the same username or email will fail.

The provenance properties offer a general strategy for debugging: (1) issue a query identifying the flow of the  undesirable behavior, and (2) provide the identified synchronizations to the LLM to reason about. In this case, given the request-id for a client experiencing the bug, we queried over the \verb|Web/request| actions bound to that specific id, and discovered the relevant flow and associated synchronizations.

% This allows lookup over all related actions, as well as which synchronizations were responsible.

The subset of synchronizations was given to an LLM, along with the following prompt: \textit{``With these syncs, when I register with a bad password, the next time I try with a good one the system complains that a user already exists. How can I fix this?''}. In response, the LLM correctly identified the problem:

\begin{small}
\begin{quote}
{\em The issue is happening because your current synchronization flow is creating the user before validating the password. When a user tries to register with an invalid password:

1. The user is created in the database via `User/register`\\
2. Then `Password/set` fails with validation error\\
3. When they try again with a valid password, the system rejects it because the username/email already exists.

To fix this, we need to validate the password before creating the user. Here's how to modify your synchronizations...}
\end{quote}
\end{small}

The LLM proposed to realign the \verb|Web/request| \\ 
with \verb|Password/validate| instead by adding a new synchronization, and modifying the existing \verb|Registration| to condition on successful validation:

\begin{small}
\begin{verbatim}
sync ValidateRegistrationPassword
when {
    Web/request: [
        method: "register" ;
        password: ?password ] 
        => [ request: ?request ]}
then { Password/validate: [ password: ?password ] }

# Only register user if password is valid
sync Registration
when {
    Web/request: [
        method: "register" ;
        username: ?username ; 
        email: ?email ;
        password: ?password ] => []
    Password/validate: [ password: ?password ] 
        => [ valid: true ] }
where { bind ( uuid() as ?user) }
then { 
    User/register: [ 
        user: ?user ; 
        name: ?username ; 
        email: ?email ] }
\end{verbatim}
\end{small}

These synchronizations, along with one additional generated synchronization for now responding on failed validation, successfully solved the issue without affecting any other existing behavior. In total, two synchronizations were added, one was altered (\verb|Registration|), while the existing behaviors of default profile creation, login token on registration, and response format continued to function as before.

This example highlights the possibility of future instrumentation that:
\begin{enumerate}
    \item upon a user marking an undesirable action, automatically marks all related actions in the flow;
    \item gathers the responsible syncs through code provenance;
    \item generates a semantic account of the issue and possible solution;
    \item forks a separate action graph to test the solution and provide the updated behavior by replaying the same inputs/outputs from the identified flow.
\end{enumerate}

%\section{Discussion}\label{discussion-section}
%This section looks to the past for precedents; to the present to candidly evaluate the current research and its successes and limitations; and to the future for the prospects it offers.

\section{Related Work}

Improving modularity (and trying to explain exactly what it means) has been a focus since the earliest days of programming. David Parnas \cite{parnas-criteria} coined the term ``information hiding'' for the criterion that each module should be ``characterized by knowledge of a design decision.'' As an example, he recommended encapsulating the definition of a data structure and its access methods, which eventually led to the notion of data abstraction and representation independence.
%(with the meaning of "information hiding" being narrowed accordingly).

A few years later, and seemingly unaware of Parnas's earlier work, Stevens, Myers and Constantine proposed ``coupling and cohesion'' as general principles for modularization \cite{stevens-myers-constantine}, which became a key part of Yourdon and Constantine's structured design \cite{yourdon-constantine}. Concepts can be viewed as an attempt not only to minimize coupling (by disallowing all references between concepts) but also to maximize cohesion, by colocating functions associated with a purpose in the world. In structured design itself, cohesion was determined instead by control and dataflow patterns in the code.

Parnas subsequently clarified the idea of coupling by articulating the idea of module dependencies (which he called the ``uses relation'') \cite{parnas-ease}, and formulated a powerful principle: that a module \textit{A} should only use a module \textit{B} if in so doing \textit{A} is easier to implement, and there is no useful subset that contains \textit{A} but not \textit{B}. As noted in Jackson's book \cite{eos} (pp. 274--276), this principle is systematically violated by the most common object-oriented programming practices.

Modularity has been heralded as the key to design in other disciplines too, notably in the use of the design structure matrix for optimizing processes \cite{eppinger-dsm} and in arguments that modularity brings economic advantages as options that can be exercised as designs evolve \cite{baldwin-clark}.

Concepts are similar to microservices \cite{fowler-microservices} in spirit although in practice microservices are not kept independent and often end up in a tangle of dependencies. The ``aggregates'' of domain driven design \cite{evans-ddd-book} also share the goal of encompassing the functionality relevant to a purpose, but do not usually form modules.

Many modularity schemes can be viewed as attempts to realize Dijkstra's notion of ``separation of concerns'' \cite{ewd-soc}. Views and viewpoints offered a separation of concerns in specifications \cite{jackson-z-views, ainsworth-viewpoint-specification, derrick-cross-viewpoint, finkelstein-viewpoints}. The tendency of object-oriented programming to conflate concerns has been widely recognized, and many remedies have been proposed \cite{ooram, aop, sop}. The aspects of aspect-oriented programming (AOP) are perhaps closest to concepts, but aspects are augmentations of a base program rather than independent modules. Moreover, aspects are not used to encapsulate all concerns, but only those that are ``cross cutting,'' which means not readily encapsulated by standard object-oriented mechanisms.

A synchronization is a kind of mediator \cite{sullivan-mediators}, a mechanism that allows functional units to be completely decoupled from one another, despite data and control flows between them at runtime. In Sullivan's scheme \cite{sullivan-mediators}, the functions that form the API of a functional unit are distinct from the events that it can signal; the mediator calls functions in response to event occurrences. Concepts, in contrast, have only actions, which act as both API functions and events.

\section{Future Prospects}

By structuring code with concepts and synchronizations, LLM-based tools can offer more than ``vibe coding'' in which results are unpredictable, limits of complexity are easily reached, and each new coding step risks undermining previous ones. 

% A variety of other potential advantages beckons. Because the synchronization engine sees all action occurrences and their causal relations, it could easily track the provenance of actions at runtime, providing valuable data for auditing and for diagnosing problems. The granularity of the program components---concepts and synchronizations---suggests another kind of provenance tracking, but at development time, with versions of each component represented as nodes in a evolving graph, along with the prompts (aka specs) used to generate them (whether by human or LLM coder).

We have seen that synchronizations can be used to factor out error handling and also the presentation/packaging of response data. Another possibility (which the authors have successfully prototyped) is to factor out the choice of persistent storage. The actions of each concept take an additional argument and result corresponding to the state before and after execution, and become pure functions. A database concept that is responsible for storing state persistently, and which hides the design secret of the choice of database system, can then be synchronized with these functional concepts. A concept's actions can even be written using updates on conventional JavaScript objects as if they were mutable, using a framework such as Immer.js \cite{immerjs}.

Distributed implementations might be achieved by allowing concept instances to run on different servers, with synchronizations as the mechanism to keep servers in step. A weakening of the synchronization semantics might allow eventual consistency. Sharding might be handled by multiple instances of a concept, each covering a set of domain objects (holding, for example, the passwords of a subset of users) with synchronizations used to determine which sharded concept instance to send an action to.

A less technically (but more socially) challenging advance that might radically alter the development of software would be the development of concept catalogs: libraries of concept designs and implementations for particular domains. Concepts within the catalog might be reviewed or even verified for correctness, so that application development becomes primarily the creation of suitable synchronizations for existing concepts. In this world, the concepts form a new kind of high-level programming language, with their actions as operations; synchronizations are then the programs written in this language.

% \section{Framing}

% \subsection{...in discussion}

% \begin{enumerate}
%     \item \textbf{Agency of scope} \textemdash{} \textit{Integrity} through guaranteeing that the only form of modification are through sets of synchronizations, where agents need not know (and cannot affect) underlying concept implementations or other synchronizations
%     \item \textbf{Agency of expressivity} \textemdash{} \textit{Granularity} through enabling the precise specification of just the concepts whose behavior you care about, including polymorphism within actions of irrelevant parameters
%     \item \textbf{Agency of information} \textemdash{} \textit{Provenance} through reifying the entire history of execution, enabling both human and LLMs to reason over runtime execution and the responsible code (syncs) for every action
% \end{enumerate}

\begin{acks}
This research benefited greatly from prior work by Abutalib Namazov, who built the first prototype scheme for generating code using concepts with an LLM; from concept coding experiments by Jennifer Lawrence; from very helpful guidance on framing and clarity by the anonymous reviewers; and from discussions with Geoffrey Litt, Kevin Sullivan, Nick Phair, and John David Nurme. 

This work was partially funded by the Machine Learning Applications (MLA) Initiative of CSAIL Alliances managed by the Computer Science and Artificial Intelligence Lab of MIT. The initiative board at the time of funding consisted of British Telecom, Cisco, and Ernst and Young.
\end{acks}

\bibliographystyle{ACM-Reference-Format}
\bibliography{refs}

% \newpage
\appendix

\section{Further Implementation Details}\label{further-details}

\subsection{Actions as State}
At the heart of the architecture is the idea that a system is composed of nothing more than data and agents. This means that in addition to data representing state, all actions associated with changes in state are fully reified as data as well. To provide maximum compatibility and to leverage existing tools, our implementation complies with Semantic Web \cite{semantic} and Linked-Data \cite{linked-data} standards, and all data is represented in terms of RDF \cite{rdf}. For example, the construction of a full action node representing a completion can uses the following JavaScript template:

\begin{small}
\begin{verbatim}
`PREFIX : <${actionSchemaIRI}>
<${action}> 
    :actions <${action}> ;
    :concept <${conceptIRI}> ;
    :name "${name}" ;
    :input ${input} ;
    :output ${output} ;
    :flow ${flow} . `
\end{verbatim}
\end{small}
This is formatted as Turtle (1.1) \cite{turtle}, where the \verb|${}| template parameters programmatically insert the corresponding bindings in code such as \verb|name| for the action name and \verb|input| for a string representation of all the input arguments (using blank node syntax for nesting).

In Turtle, triples ending with a period \verb|.| represent an edge (property) between two nodes (a subject and object), while the semicolon \verb|;| is a shorthand for describing multiple properties of the same subject. This says that every instance of an action is associated with a concept, its name, and more data in the graph about its inputs and outputs, with the \verb|:flow| parameter providing runtime provenance as described in Section \ref{flow-scope}.

\subsection{State and Information} \label{state-information}

Each concept is associated with its own \textit{named graph}, a unique identifier for a logical grouping of data
%representing its information
, which can both be versioned and shared to describe a particular instance for a given application. Every application and its full action history are then represented as another named graph, which can also be versioned and staged: by referencing \verb|<http://example.com/App/dev>| as opposed to\\ \verb|<http://example.com/App/stage>|, eg, an entire execution environment can be separated cleanly at the storage level.

System state can be stored in any graph-based medium, which includes traditional relational/SQL databases and document stores in in addition to dedicated graph databases, each with its own performance characteristics. For our implementation, we chose the N3.js \cite{n3} low-level library for an in-memory RDF compatible graph store, and the Comunica \cite{comunica} SPARQL query engine to perform queries.

\subsection{Computational Agents}

Concepts and synchronizations alike are represented uniformly in our implementation by
%Instead of specifying any particular framework or engine that must handle computation, all state evolves purely by the behavior of 
\textit{computational agents} (similar to actors \cite{actor}, but simpler in that they cannot spawn other agents or pass messages to one another).
%these agents are simpler, cannot spawn other agents, and cannot directly pass messages to one another. They are described as computational agents because they have only one form of agency: the freedom to choose how and when to perform a computation. 

%A concept is implemented by a physical storage of graph-like data, and a computational agent responsible for implementing its behavior. 

%Computational agents are entirely characterized by their \textit{obligations}, which are the conditions under which they should compute.

%This system is comprised of the following computational agents: 1 for each concept, and 1 for synchronizations.

\subsubsection{Concept Agents}

A concept is implemented by a physical storage of graph-like data, and a computational agent responsible for implementing its behavior, which is described by a SPARQL query with this template:

%As all data exists in a graph, the obligations of a concept agent can be described by the following template for a SPARQL query:

\begin{small}
\begin{verbatim}
`PREFIX : <${actionSchemaIRI}>
CONSTRUCT {
    ?_action :input ${input}
}
FROM <${actionGraphIRI}>
WHERE {
    ?_action 
        :actions ?_action ;
        :concept <${conceptIRI}> ;
        :name "${name}" ;
        :input ${input} .
    FILTER NOT EXISTS { ?_action :output [] }
}`
\end{verbatim}
\end{small}

The \verb|CONSTRUCT| keyword specifies that the query should return the graph specified, which in this case includes any arbitrarily nested shape of nodes described by the \verb|${input}| template parameter, and which is also fully specified in the action signature of the associated concept. This query says to simply find all actions
%\verb|?_action|s 
that do not yet have an output
%\verb|:output|, 
and capture all the inputs for processing.
%\verb|:input|. 

The behavior of a concept agent thus involves these steps:

\begin{enumerate}
    \item Query the action graph (as above) for outstanding actions associated with the concept;
    \item Perform each action and compute the state changes to insert into the concept state graph;
    \item Insert the completion record with the output into the action graph.
\end{enumerate}

The behavior of a concept agent might even be thought of as a kind of reverse synchronization that goes from invocation to completion, with
\verb|ACTION()| representing the arbitrary computational function of the agent:

\begin{small}
\begin{verbatim}
when { Concept/action [ :input ?input ] }
where { bind ( ACTION(?input) as ?output) }
then { Concept/action [ :input ?input ] 
                    => [ :output ?output ] }
\end{verbatim}
\end{small}

%This is a synchronization "in reverse", as it goes from invocation to completion, but the syntax reifies the idea of what is going on when a machine fulfills a concept action.

\subsubsection{Synchronization Agents} \label{sync-agent}

Synchronization agents function in a similar way: they look at outstanding action \textit{completions} that have not yet been synchronized according to the specification, and insert \textit{invocations} with the appropriately bound inputs. Their behavior can be characterized entirely by a transactional update query of the form:

\begin{small}
\begin{verbatim}
`...prefixes...
INSERT {
    ?_then_1 
        :actions ?_then_1 ;
         ...properties... ;
        :flow ?_flow .
    ...thens...
    ?_when_1 :${syncName} ?_then_1 .
    ...synced...
}
WHERE {
    ...wheres...
    ?_when_1
        :actions ?_when_1 ;
        ...properties... ;
        :flow ?_flow .
    ...whens...
    FILTER NOT EXISTS { ?_when_1 :${syncName} [] }
    ...filters...
}`
\end{verbatim}
\end{small}

The triple dots \verb|...thens...| indicate elided repetitions of (for example) other possible actions specified in the \verb|then| clause. The shared \verb|?_flow| token and \verb|FILTER NOT EXISTS| for the synchronization edge represent the implementations for these ideas described in Sections \ref{flow-scope} and \ref{provenance-firing}. 

% \begin{enumerate}
%     \item \textbf{Maintain flow.} All actions in the \verb|when| clause must be associated with the same \verb|?_flow|, and all actions of the the \verb|then| clause to be inserted must also be colored with the same \verb|?_flow| identifier. This allows reasoning over causally related actions: that the same \verb|User/register| resulting from a \verb|Web/request| is described, and not any other user registrations that may occur in a given timeframe.
%     \item \textbf{Sync tracking.} All resulting \verb|then|s are pair-wise associated with all causal \verb|when|s by an edge labeled specifically with the \verb|${syncName}|, which prevents infinite re-firing of action invocations that would otherwise match completions. 
% \end{enumerate}

\section{Concept Specification Examples}\label{concept-spec-appendix}

Each concept specification was generated using a system prompt that explained the overall pattern and the details of the specification language. The \texttt{User} concept specification below 
%in section \ref{concept-spec-user} 
was given as an example.
%of a proper specification in the system prompt, which walks through each section and keyword in a similar manner to section \ref{concept-language-section}. 

\subsection{User: sample spec used in prompt}\label{concept-spec-user}
\begin{small}
\begin{verbatim}
concept User [U]
purpose 
    to associate identifying information with users
state
    users: set U
    name: U -> string
    email: U -> string
actions
    register [ 
        user: U ; 
        name: string ; 
        email: string ] => [ user: U ]
        associate user with users
        associate name and email unique + valid
        return the user reference
    register [ 
        user: U ; 
        name: string ; 
        email: string ] => [ error: string ]
        if either name/email is invalid or !unique
        return the error description
    update [ user: U ; name: string ] => [ user: U ]
        if name is unique, update user's name
        return the user reference
    update [ user: U ; name: string ] 
        => [ error: string ]
        if name is not-unique, describe error
        return the error description
    update [ user: U ; email: string ] 
        => [ user: U ]
        if email is unique + valid, update id's email
        return the user reference
    update [ user: U ; email: string ] 
        => [ error: string ]
        if email is not-unique or invalid
        return the error description
operational principle
    after register [] => [ user: x ] 
    and update [ name: "xavier" ] => [ user: x ]
    for any ?u such that ?u.name = "xavier", ?u = x
\end{verbatim}    
\end{small}
\subsection{Profile: a generated concept}\label{concept-spec-profile}

The prompt for the \verb|Profile| concept was:

\begin{quote}{\em
%\texttt{
Create a concept for Profile, where the purpose is to associate descriptive information with an individual. A profile consists of a bio and an image (can be a string storing a URL).  
%}
}
\end{quote}

which generated the following specification:

\begin{small}
\begin{verbatim}
concept Profile [P, U]
purpose 
    to associate descriptive information with users
state
    profiles: set P
    profile: U -> P
    bio: P -> string
    image: P -> string
actions
    register [ profile: P ; user: U ] 
        => [ profile: P ]
        add profile to profiles
        associate user with profile
        add a default blank bio and image to profile
        return profile
    update [ profile: P ; bio: string ] 
        => [ profile: P ]
        update profile with bio
        return profile
    update [ profile: P ; image: string ] 
        => [ profile: P ]
        if image is valid (URL, base64, etc.)
        update profile with image
        return profile
    update [ profile: P ; image: string ] 
        => [ error: string ]
        if image is invalid, describe error
        return error
operational principle
    after register [ profile: p ; user: u ] 
    => [ profile: p ] 
    and update [ profile: p ; bio: "Hello world" ] 
    => [ profile: p ]
    and update [ profile: p ; image: "pic.jpg" ]
    => [ profile: p ]
    the profile p contains both the updated bio 
    "Hello world" and original image "pic.jpg"
\end{verbatim}    
\end{small}

\section{Synchronization Generation}

The general structure of a synchronization prompt included (1) a system prompt explaining the key ideas (an earlier version of Section \ref{sync-language-section}), (2) the set of relevant concepts in only their specification form (no code), and (3) a user prompt describing the desired behavior. The following example shows how synchronizations were added to accommodate new concepts.
%could be extended simultaneously with synchronization generation. 
% After completing the \verb|User| registration flow incrementally, testing synchronizations implementing functionality related to \verb|User|, \verb|Profile|, \verb|Password| and \verb|JwT|, and then 

After generating the \verb|Article| concept, this prompt was issued (along with copy-pasted RealWorld documentation for endpoint formats, to support the ad hoc response synchronization):
%to generate functionality for \verb|Article|.

%At the same time, the RealWorld specification requires both tagging and favoriting to feature in its article endpoints that conflate these concepts. In addition to copy-pasting the RealWorld documentation for endpoints, the prompt was:

%\texttt{
\begin{quote}{\em 
Given the RealWorld specification, create a set of syncs for basic article management. You may propose a Tag concept and Favorite concept to help. }
\end{quote}

The LLM responded with the \verb|Tag| and \verb|Favorite| concept specifications, which could then later have implementations generated. These concept specifications did not pose issues and were generated in one-shot, and the generated implementations were sufficient for the purposes of synchronization with \verb|Article|. In addition, the following is an excerpt of the synchronizations generated for article creation, along with a few demonstrations of error handling, authentication, and associating tags. 

\begin{small}
\begin{verbatim}
sync CreateArticle
when {
    Web/request: [
        method: "create_article" ;
        title: ?title ;
        description: ?description ;
        body: ?body ;
        token: ?token ] 
        => [ request: ?request ] }
then { 
    JWT/verify: [ 
        token: ?token ] }

sync HandleCreateArticleAuth
when {
    Web/request: [
        method: "create_article" ] 
        => [ request: ?request ]
    JWT/verify: [] 
        => [ error: ?error ] }
then { 
    Web/respond: [ 
        request: ?request ; 
        error: ?error ; 
        code: 401 ] }

sync PerformCreateArticle
when {
    Web/request: [
        method: "create_article" ;
        title: ?title ;
        description: ?description ;
        body: ?body ] 
        => []
    JWT/verify: [] 
        => [ user: ?user ] }
where {
    BIND ( UUID() as ?article ) }
then { 
    Article/create: [ 
        article: ?article ; 
        title: ?title ; 
        description: ?description ; 
        body: ?body ; 
        author: ?user ] }

sync HandleCreateArticleError
when {
    Web/request: [
        method: "create_article" ] 
        => [ request: ?request ]
    Article/create: [] 
        => [ error: ?error ] }
then { 
    Web/respond: [ 
        request: ?request ; 
        error: ?error ; 
        code: 422 ] }

sync CreateArticleFormat
when {
    Web/request: [
        method: "create_article" ] 
        => [ request: ?request ]
    Article/create: [] 
        => [ article: ?article ]
    JWT/verify: [] 
        => [ user: ?currentUser ] }
then {
    Web/format: [
        type: "article_auth" ;
        article: ?article ;
        currentUser: ?currentUser ;
        request: ?request ] }

sync AddTagsToArticle
when {
    Web/request: [
        method: "create_article" ;
        tagList: ?tag ] 
        => []
    Article/create: [] 
        => [ slug: ?slug ] }
then { 
    Tag/add: [ target: ?slug ; tag: ?tag ] }

\end{verbatim}
\end{small}

\subsection{Response Formatting}
The \verb|Web/format| action called by the previous synchronizations factors out application-specific formatting details. In RealWorld, web responses are highly entangled, involving information from multiple concepts. The following example of 
% the unauthorized version of 
a formatted article response shows how the language handles this complexity, leveraging SPARQL built-ins such as OPTIONAL graph joins for conditionally including information from other concepts. 

%One challenge with constructing elaborate data structures (such as nested response formats) is the aggregation of results.
As explained in Section~\ref{where-section}, the \verb|where| clause is a declarative query (akin to a SPARQL SELECT) whose result is a set of ``frames'' (each comprising a set of bindings of variables to values). This means that an article with multiple tags would result in a separate frame for each tag. To aggregate these by article, invoking the \verb|then| clause once for each article and the associated values for other variables, 
%on each article in at least as many such tuples, or "frames" of bindings. For the purposes of aggregating these results for each \verb|then| clause that formulates action invocations, 
we use a special variable \verb|?_eachthen|. This mechanism functions identically to SPARQL's GROUP BY for the purposes of specifying values to aggregate over (e.g. allowing grouping by composite keys), but circumvents the restriction that selections must be limited to aggregates and constants.
%, which groups results by whatever is bound to it.

\begin{small}
\begin{verbatim}
sync FormatArticle
when {
    Web/format: [
        type: "article" ;
        article: ?article ;
        request: ?request ] 
        => [] }
where {
    Article: { 
        ?article title: ?title ; 
        description: ?description ; 
        body: ?body ;
        slug: ?slug ;
        createdAt: ?createdAt ;
        updatedAt: ?updatedAt ;
        author: ?author }
    User: { ?author name: ?authorName }
    Profile: {
        ?author profile: ?profile .
        ?profile bio: ?authorBio ;
        image: ?authorImage }
        
    # Get tags for this article if they exist
    OPTIONAL {
        Tag: { ?article tag: ?tag } }

    # Get favorites count if it exists
    OPTIONAL {
        Favorite: { ?article count: ?count } }
    
    # Aggregate all results by unique article ID
    BIND ( ?article AS ?_eachthen ) }
then {
    Web/respond: [ 
        request: ?request ;
        body: [ 
            article: [
                slug: ?slug ;
                title: ?title ;
                description: ?description ;
                body: ?body ;
                tagList: ?tag ;
                createdAt: ?createdAt ;
                updatedAt: ?updatedAt ;
                favorited: false ;
                favoritesCount: ?count ;
                author: [
                    username: ?authorName ;
                    bio: ?authorBio ;
                    image: ?authorImage ;
                    following: false ] ] ] ] }
\end{verbatim}
\end{small}
One final complication: if the article has no tag, the \verb|tagList| field would not appear in the response object at all, but the RealWorld API would require a field with an empty list. To fix this, we can bind the field to a variable \verb|?tagList| with:
\begin{small}
\begin{verbatim}
  BIND ( COALESCE( ?tag , rdf:nil ) AS ?tagList )
\end{verbatim}
\end{small}
\end{document}